# Macro-particle FEL model with self-consistent spontaneous radiation[*]

Vladimir N. Litvinenko

*Department of Physics and Astronomy, Stony Brook University, Stony Brook, USA*

**Abstract**

Spontaneous radiation plays an important role in SASE FELs and storage ring FELs operating in giant pulse mode. It defines the correlation function of the FEL radiation as well as its many spectral features. Simulations of these systems using randomly distributed macro-particles with charge much higher that of a single electron create the problem of anomalously strong spontaneous radiation, limiting the capabilities of many FEL codes.

In this paper we present a self-consistent macro-particle model which provided statistically exact simulation of multi-mode, multi-harmonic and multi-frequency short-wavelength 3-D FELs including the high power and saturation effects. The use of macro-particle clones allows both spontaneous and induced radiation to be treated in the same fashion. Simulations using this model do not require a seed and provide complete temporal and spatial structure of the FEL optical field.

## 1. Introduction.

The idea of this paper is based on a simple observation that a number of processes (such as an FEL gain or an electron cooling force) are quadratic with respect to the carrier charge. This simple observation provides for a simple use of marc-particles and clones to separate spontaneous effects (noise) from collectives ones.

This paper was written and was presented at FEL 2002 conference [1]. By a technical reason the paper was not published, but the idea had been successfully used and tested for various applications from FELs to electron cooling [2-6]. During twelve years after the presentation of this paper the progress in computer simulations and available computer power increased dramatically and many systems can be simulated with real number of electrons. Nevertheless, the proposed idea still can be useful for simulating processes with very large number of particles or shot noise, or simply to speed-up simulations. This is the reason for this publication.

A number of complex processes in modern SASE and oscillator FELs [7-11] are properly described only by multi-dimensional time-dependent FEL codes (see for example [12-16]). Self-consistent treatment of spontaneous radiation is of the foremost importance for the attainment of correct temporal and spectral features of FEL pulses.

Even though the power and memory of modern computers have improved, it is still a challenge to simulate a 3-D FEL with $10^9$-$10^{11}$ real electrons, and in the case of storage ring

---

[*] This work is supported by the Dean of Natural Sciences, Duke University

FELs [15] to do it for $10^3$-$10^7$ passes. Further, if such code would exist, its usefulness for the optimization of FEL design and for theoretical studies will very time consuming if even possible. Therefore, most of modern multi-dimensional FEL codes are based on some kind of the macro-particle model. Each macro-particle comprises a large number of electrons, which is naturally creates the problem of proper treatment of the spontaneous radiation. Modern multi-dimensional time-dependent codes use either an exact analytical model as fully 3D-code #felTD/#uvfel [15], or a special seeding of the macro-particles as in the 2-D FEL code GINGER [13]. The use of self-consistent analytical approach is limited to the case of a low gain FEL [15,17] and can not be used for high gain and SASE FELs. Generally speaking, the high gain and SASE FELs require treating both induced and spontaneous radiation in the same fashion. A detailed extent of the problems related to proper seeding of macro-particles is elegantly described in recent paper by Fawley [18].

In a typical approach, thousands to millions of individual electrons are represented by a macro-particle with Dirac's δ-function distribution in 6-D phase space The specific choice of canonical coordinates and momenta is of no importance, thus we have :

$$f(q_i, p_i, z) = \prod_{i=1}^{3} \delta(q_i - q_{ia}) \prod_{i=1}^{3} \delta(p_i - p_{ia}); \quad (1)$$

The charge of macro-particle is much larger than electron charge $e$:

$$q_{mp} = n_{mp} e; \quad n_{mp} = \frac{N_e}{N_{mp}} >>> 1; \quad (2)$$

where and $n_{mp}$ is the number of electrons per macro-particle (not necessary integer), $N_e$ is number of electrons in the beam, $N_{pm}$ is the number of macro-particles. A simple minded use of randomly spread macro-particles will result in an artificial increase in shot noise by a factor of $\sqrt{n_{mp}}$ and correspondingly in $n_{mp}$ times higher spontaneous radiation power.

The most common approach for solving this problem is seeding a subset of 2M macro-particles with even time intervals to eliminate the artificial shot noise [18]:

$$\Delta t = \frac{1}{2M} \cdot \frac{\lambda_o}{c}; \quad (3)$$

where $\lambda_o$ is the fundamental wavelength of radiation, and $c$ is the speed of light. This method eliminates short noise up to the $M^{th}$ harmonic of the fundamental wavelength. As described in [18], by adding proper random noise to these time intervals, one can generate an ensemble of macro-particles with proper statistical properties of the simulated spontaneous radiation. This approach is well tested and proven to work reliably for reproducing on-axis radiation correctly. Nevertheless, this approach

- Requires prior knowledge of the wavelength where lasing will occur.
- Has an intrinsic FEL wave-packet bandwidth limitation about $\Delta\lambda \propto \lambda_o / \sqrt{n_{mp}}$ because of violation of the artificial short noise cancellation. Using $M$ evenly distributed macro-particles (as in eq(3)) gives a non-zero short noise at wavelength $\lambda$ different from the fundamental wavelength $\lambda_o$ or its harmonic:

$$sn = \frac{n_{mp}}{2M} \sum_{m=0}^{2M-1} e^{i\pi n(1+\varepsilon)/M} = \frac{n_{mp}}{2M} \frac{1-e^{2\pi i \varepsilon}}{1-e^{i\pi(1+\varepsilon)/M}}; \text{ where } \varepsilon = \frac{\lambda_o}{\lambda} - \text{integer}\left\lfloor \frac{\lambda_o}{\lambda} \right\rfloor.$$

Which of the case of $M >> 1$ and $\varepsilon << 1$ yields $sn \approx n_{mp}\varepsilon$. Combined with the random noise introduced by time intervals, this effect can double the value of shot noise at $\varepsilon \approx 1/\sqrt{n_{mp}}$.

- Requires 2M macro-particles per one 6-D coordinate to properly simulate spontaneous radiation at the $M^{th}$ harmonic of the fundamental wavelength.
- Can not be used for an arbitrarily large number of harmonics.

In this paper we present an approach which eliminates the above limitations and can be used for in many multi-dimensional time-dependent FEL codes without bandwidth limitation. The only shortcoming of this method is that it is applicable to FELs generating odd harmonics of the fundamental exclusively. This approach was developed by the author in 1993-1994 and used for a prototype of the full FEL code[1]. It is important to emphasize that this approach does not limit the FEL code to any specific model or approach and it can be incorporated into already existing codes with little effort.

## 2. Macro-particles and their clones.

The idea of using "positron" clones for macro-particles of "electrons" is based on the following observations:

1. Induced radiation in an FEL is a second order effect (for a beam without initial pre-bunching) and depends neither on the sign of the particle's charge nor its value. It depends only on $(e/m)^2$ (see for example [19-20]) and therefore substituting electrons with positrons in an FEL yields the same results for fundamental wavelength and its odd harmonics.

2. For a given trajectory, the spontaneous radiation for either the electrons or the positrons is the same with the exception of the sign of the electric and magnetic field. It is obvious that if an electron and a positron propagate along the same trajectory $\vec{r}(t)$, their radiation fields cancel.

3. The positron trajectory will be identical to that of the electron if we "switch off" the FEL interaction and reverse the sign of the magnetic field in the FEL system (including all magnetic elements: wigglers, quadrupoles, etc.).

4. If we put a macro-particle composed of $n_e$ electrons and a positron clone composed of $n_p$ positrons on this trajectory, the power of spontaneous radiation from such a system will be equivalent to that from $N = (n_e - n_p)^2$ randomly distributed electrons.

5. Randomly distributed $N_{mp}$ pairs of macro-particles and their clones with charges of:

$$Q_{mp} = -e\frac{n_{mp} + \sqrt{n_{mp}}}{2}; Q_{clone} = e\frac{n_{mp} - \sqrt{n_{mp}}}{2} \qquad (3)$$

will generate spontaneous radiation equivalent in power, spectrum and statistics to that of an ensemble of $N_e = N_{mp} \cdot n_{mp}$ randomly distributed electrons.

---

[1] This approach was used by my graduate student, B. Burnham, for FEL code described in his Ph.D. thesis (Duke University, 1995). Simulations of spontaneous radiation provided excellent agreement with theory. The part of the code related to the FEL interactions was never completely debugged is currently abandoned. Presently the author and Dr. O. Shevchenko (Novosibirsk) are developing a new generic FEL code with self-consistent 6D(beam) x 3D(EM field) treatment to be published separately.

6. Switching on the FEL interactions will separate electrons and positrons according to the standard FEL equations. This separation is responsible for the bunching (see section 4) and the induced radiation in an FEL.

It can be shown that there is no difference if we consider the positrons to be the macro-particle and the electrons to be the clone. Let's consider an ensemble of pairs (macro-particles of electrons and their positron clones) with desirable distribution in 6-D phase space given by initial conditions at the FEL entrance, and with random optical phase distribution [2]. By design, the clone and the macro-particle are initially located at exactly the same point in phase space. Without the FEL interaction, they will continue to share the same point in the 6-D phase space while traversing the FEL system. The assignment of the proper charges to the macro-particles and clones in eq.(3) produces spontaneous radiation statistically equivalent to that of the real beam[3].

By the design, this Macro-Particle/Clone pair (MP/C) model has a classical, δ-function like distribution, and completely white shot noise. This means that these pairs provide statistical equivalence to classical spontaneous radiation from an electron ensemble in the full range of the radiated spectrum. The only difference in the clone's treatment from the macro-particle being that the sign of magnetic field of the FEL system is reversed in order to make the clone follow the trajectory of the macro-particle in the absence of the FEL interaction.

The intensity of spontaneous radiation is proportional to the number of macro-particles and the square of the net "pair" charge $q = Q_{mp} + Q_{clone}$, i.e. it is sign independent (see Fig.1).

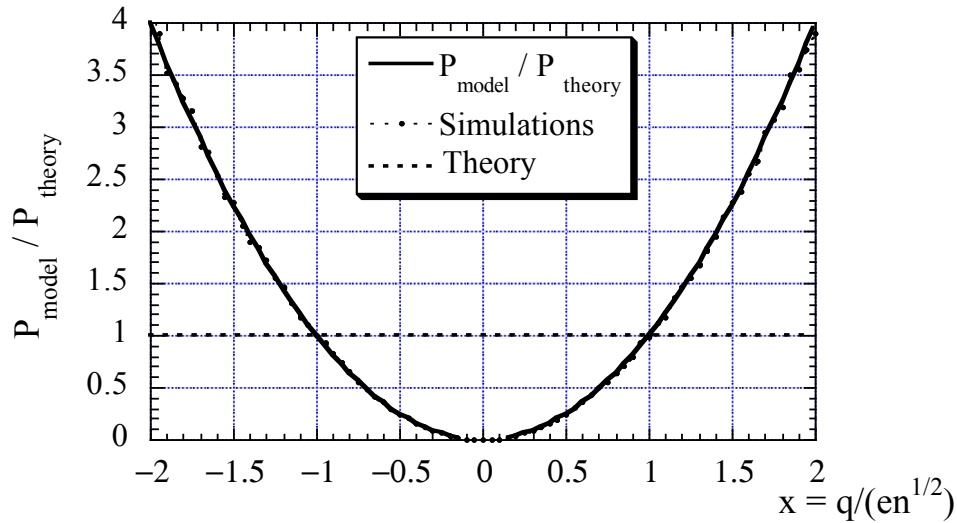

Figure 1. Power of spontaneous radiation as a function of the net charge of macro-particle/clone pair. The continuous line represents the theoretical value. Points show the results of simulations with 1,000 pairs per slippage length. The dashed line corresponds to the correct net charge setting given by eq.(3).

---

[2] I.e. random optical phase at the scale of the FEL wavelength. Requirements for statistically independent random distributions are well described in [18,21]
[3] It was noted by N.A. Vinokurov [23], that the difference between statistical properties of the real beam and the MP/C model (as well as for any macro-particle model) will appear in the second order of power correlation function.

Controlling the ratio $x = |q/e\sqrt{n_{mp}}|$ between the net "pair" charge and the natural RMS value of the short noise, one can study the influence of the intensity of spontaneous radiation on the processes in an FEL. It can be also very handy for debugging and testing FEL codes in amplifier mode with "seeded" optical input.

### 3. MP/C model: the equivalency of electrons and positrons.

The equivalency of electrons and positrons in an FEL can be proven from first principles. The equation of motion of electrons and positrons in an FEL are identical if the of the FEL magnetic field is reversed. Separating the vector potential of the EM field into its DC and AC components $\vec{A} = \vec{A}_{DC} + \vec{A}_{ac}$, flipping the sign of magnetic field in the FEL system $\vec{A}_{DC} \rightarrow -\vec{A}_{DC}$ for positrons and choosing the radiation gauge, we have [4]:

$$\varphi = 0; \ div\vec{A} = 0;$$
$$\frac{1}{c^2}\frac{\partial^2 \vec{A}_{AC}}{\partial t^2} - \vec{\nabla}^2 \vec{A}_{AC} = \frac{4\pi}{c}\vec{j}_t; \ \vec{j}(\vec{r},t) = e\sum_{part}\frac{d\vec{r}_{part}}{dt}\delta(\vec{r}-\vec{r}_{part}(t)). \quad (4)$$

The classical equations of motion are determined by the Hamiltonian [17]:

$$P = \left\{ m^2c^2 + \left(\vec{P} - \frac{e\vec{A}_{DC}}{c} - \frac{e\vec{A}_{AC}}{c}\right)^2 \right\}^{1/2}. \quad (5)$$

Switching the sign of particle's charge $e \rightarrow -e$ automatically flips the sign $\vec{A}_{AC} \rightarrow -\vec{A}_{AC}$ as follows from equation (4)[5]. The Hamiltonian (5) remains unchanged, proving the equivalence of both systems. Thus, starting from dentical initial conditions ($\vec{A}_{AC} = 0$ and the same 6-D micro-ensemble), the electrons and positrons will generate identical FEL field with opposite sign. The consequences of the sign flip are discussed further in the following section.

### 4. MP/C model: difference between the electrons and positrons.

Even though the MP/C model reproduces exact statistical features of the spontaneous radiation of the real beam exactly, there is a subtle problem in the model. The extent of this problem appears when one combines two beams together in an FEL and observes the bunching features of the electrons and positrons during FEL interactions.

Specifically, the optical wave of wavelength $\lambda_m$ co-propagating with MP/C pairs in an FEL caused the bunching positrons with π-phase shift compared with bunching of electrons as shown in Fig.2 and Fig.3. Because of the sign difference in charge, the amplitudes of the density modulation occurring at all odd harmonics of $\lambda_m$ ($\lambda_{odd} = \lambda_m /(2n+1)$) are identical for e$^+$ and e$^-$ beams. In contrast, all the amplitudes of even harmonics ($\lambda_{even} = \lambda_m /2n$) have the same absolute value but opposite signs. In other words, the current modulation at all *even harmonics* is *artificially suppressed* by a factor of $1/\sqrt{n_{mp}}$ in the MP/C model. We have not found a simple cure for this problem as yet.

---

[4] See discussion of the space-charge effect in the next section.
[5] Assuming the motion of particles remains unchanged

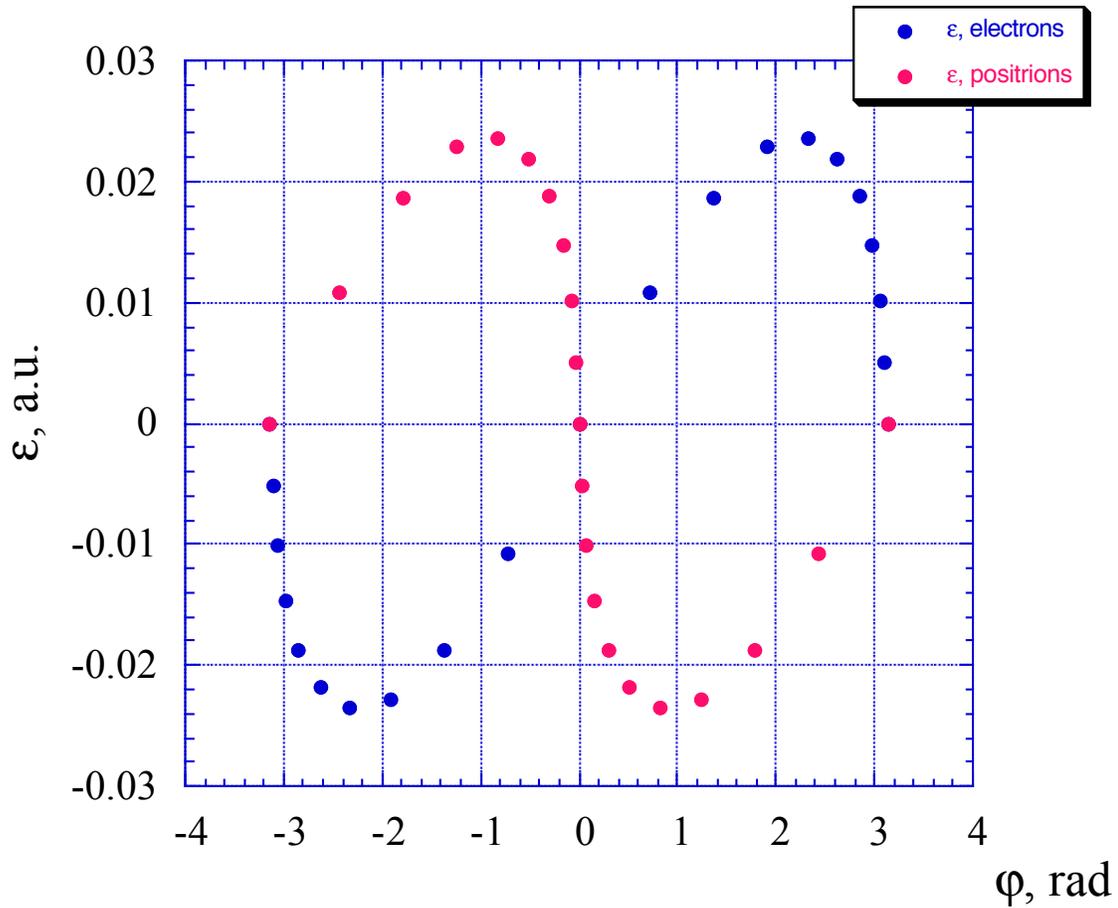

Figure 2. Micro-bunching of electrons and positrons in an FEL model ($N_w$=30) caused by an external laser field is identical with the exception of the $\pi$ phase shift. Only one beam-let is shown.

Therefore, the MP/C model in its present shape can not be used for simulation of FELs where coherent even harmonics[6] play a significant role. Examples of such systems include an FEL with an asymmetric wiggler producing strong even harmonic content in the field or an FEL with gain-length comparable to its wiggler period[7].

This limitation makes the MP/C model applicable only for FELs where coherent even harmonic radiation is very weak and its influence on the FEL process can be neglected. Otherwise, even coherent harmonics can be extracted from MP/C model by a $\sqrt{n_{mp}}$ scaling.

Fortunately for the model, most FELs (including short wavelength SASE FELs) are based on planar or helical wigglers, which produce only odd harmonics of the fundamental period and have gain lengths many rimes their wiggler period. In these cases, only odd harmonics are present in the on-axis radiation [19]. In such FEL systems, even harmonics are radiated off-axis and are of much lower power than the odd harmonics.

---

[6] Notice that spontaneous radiation at even harmonics will not have such a problem – the $\delta$-function like short noise takes care of this effect.
[7] Experimental results from VISA FEL, which has world's shortest FEL gain length [26], indicate strong presence of second coherent harmonic in the spectrum.

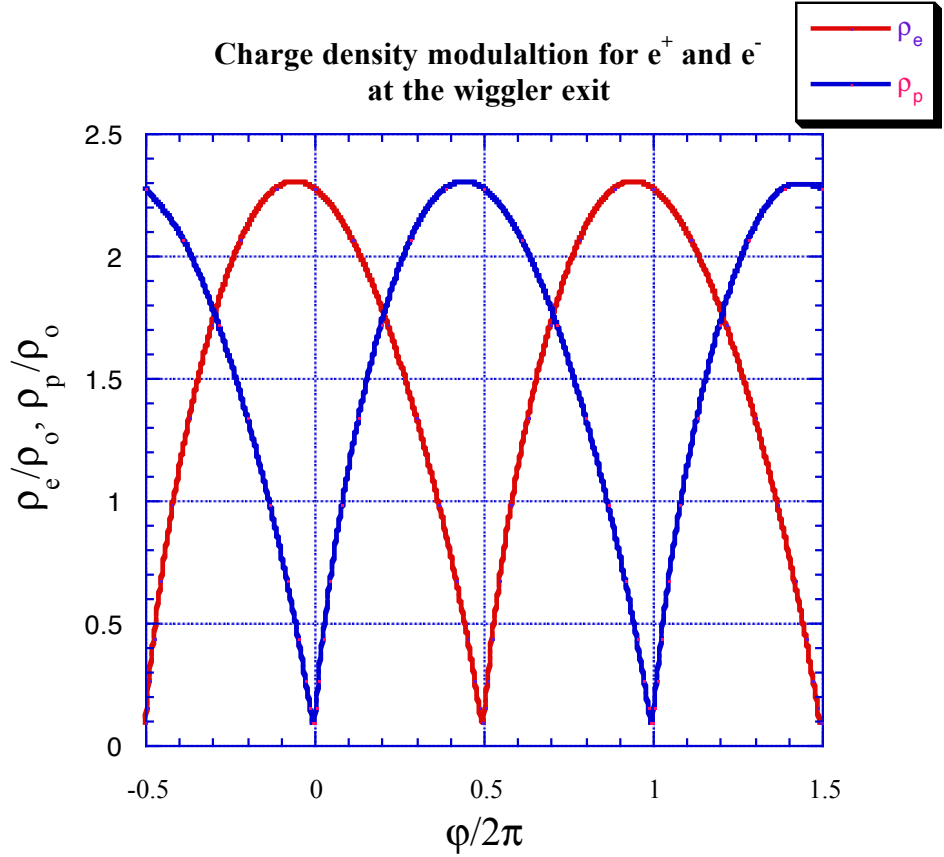

Figure 3. Micro-bunching of electrons and positrons in FEL model (the same as in Fig. 2). The amplitudes of the harmonics in the density modulation are identical in the amplitude. The π-phase shift provides for the same sign of the density modulation on all odd harmonics and opposite sing of all even harmonics.

The radiation of even harmonics is concentrated at angles of $\theta \propto 1/\gamma$, where $\gamma$ is the relativistic factor of the electrons. This angle is normally much larger than the divergence of the FEL optical beam $\theta_o \propto 1/\gamma\sqrt{N_w}$, where $N_w$ is the number of wiggler periods per gain length.

## 5. Applicability.

In conclusion, we would like to emphasize the simplicity of this model, which can be used for any type of 6-D phase space distribution (see Fig.4). The standard requirements for number of macro-particles per slippage or gain length, and per transverse coherence volume [23,18] are directly applicable to the MP/C model [8].

---
[8] Experience with #fel3D/#uvfel simulating the Super-pulses in the OK-4/Duke storage ring FEL demonstrated better performance when 1,000,000 (~1,000-3,000 MPs/slipage length) MPs. When compared with previously used

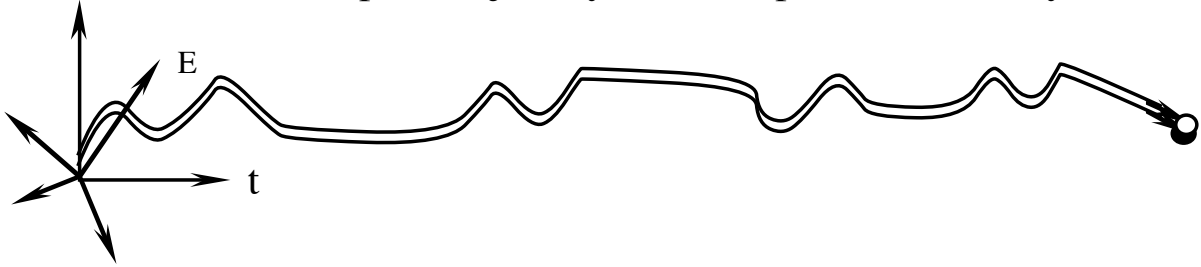

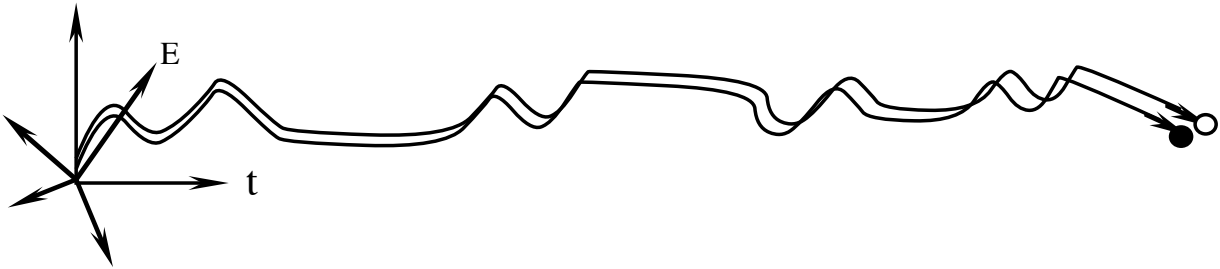

Figure 4. Schematic of macro-particle and clone trajectories without (top) and with FEL interaction (bottom). Without FEL interaction the MP/C pair occupies the same point in 6D-phase space. The FEL interaction tears clone ○ apart from macro-particle ●, predominantly in the "energy-time" plane of 6D phase space.

Most operational FEL codes can use the MP/C model with only modest modifications: in the FEL the force acting on a positron from the DC magnetic field must be the same if it were applied to an electron located at the same point in the phase space. It is important to note that the MP/C model is generally compatible with programs including space charge effects and Coherent Synchrotron Radiation (CSR). In classical statistical systems, the existence of the time scale τ much longer than the optical period and much shorter than the electron bunch duration provides for clear separation of microscopic (AC) and macroscopic fields. This time scale clearly exists for short wavelength FELs with e-bunch duration from 0.1 to 100 psec and optical period from 0.001 to 0.000001 psec. All DC-like effects (wake-fields including CSR and space charge fields) should be calculated for the electron beam and applied to the clones with reversed sign in the same fashion as was done with magnetic field of FEL system.

## 6. Acknowledgements .

The author would like to thank Nikolay Vinokurov and Oleg Shevchenko (BINP, Novosibirsk, Russia) for fruitful discussions of the MP/C concept and its limitations, Emily Longhi (Diamond Light Source, UK) for important suggestions and help in preparation of this paper.

---

30,000 MPs (~30-100 MPs/slipage length), the noise in the FEL gain and the wave-packet structure reduced roughly by a factor 10 [24].

I want to thanks my colleagues from Tech X and RadiaSoft for reminding me that this paper, being distributed privately for a dozen of years, remained unpublished.## References

1. V.N Litvinenko, Macro-particle FEL model with self-consistent spontaneous radiation, presented at FEL 2002 conference, WEP45, September 9-13, 2002, Argonne, IL, USA. http://www.aps.anl.gov/News/Conferences/2002/fel2002/PosterAbs.pdf
2. Modulator simulations for coherent electron cooling using a variable density electron beam, George I. Bell, Ilya Pogorelov, Brian T. Schwartz, David L. Bruhwiler, Vladimir Litvinenko, Gang Wang, Yue Hao, arXiv:1404.2320, 8 Apr 2014, http://arxiv.org/pdf/1404.2320.pdf
3. W.M. Fawley, Some Issues and Subtleties in Numerical Simulation of X-ray FEL's, http://www.osti.gov/scitech/servlets/purl/805131
4. David L. Bruhwiler, Simulating Single-pass Dynamics for Relativistic Electron Cooling, Beam Dynamics Newsletter No. 65, 2014, p. 117, http://www-bd.fnal.gov/icfabd/Newsletter65.pdf
5. Scientific Assessment of High-Power Free-Electron Laser Technology (2009), Committee on a Scientific Assessment of Free-Electron Laser Technology for Naval Applications, NATIONAL RESEARCH COUNCIL OF THE NATIONAL ACADEMIES, Division on Engineering and Physical Sciences, http://www.nap.edu/openbook.php?record_id=12484&page=14
6. L. van der Meer, private communication
7. J. Rossbach et al, Nucl. Instr. and Meth. **A475**, (2001) 13
8. N.D. Arnold et al, Nucl. Instr. and Meth. **A475** (2001) 20
9. J. Wu, L.H. Yu et al, Nucl. Instr. and Meth. **A475** (2001) 104
10. I.V. Pinayev et al., Nucl. Instr. and Meth. **A475** (2001) 222
11. V.N. Litvinenko "Duke XUV Storage Ring FEL and Coherent X-Ray Harmonics Generation", In Proc. of International Workshop on Generation and Application of Coherent X-Rays, KEK, Tsukuba, Japan, February 29-March 1, 1996, p. 282
12. R.J.Dejus, O.A.Shevchenko, N.A.Vinokurov, Nucl. Instr. and Meth **A445** (2000) 19
13. W.M. Fawley, LBNL report LBNL-49625 (2002)
14. S. Reiche, Nucl. Instr. and Meth. **A429** (1999) 243
15. V.N. Litvinenko et al., Nucl. Instr. and Meth. **A475** (2001) 65, V.N.Litvinenko, B.Burnham, J.M.J. Madey, Y.Wu, Nucl. Instr. and Meth. **A358** (1995) 334, V.N.Litvinenko, B.Burnham, J.M.J.Madey, Y.Wu, SPIE v. 2521 (1995) 79
16. E.L. Saldin, E.A. Scheidmiller, M.V. Yurkov, Nucl. Instr. and Meth. **A429** (1999) 233
17. V.N. Litvinenko, Nucl. Instr. and Meth. **A359** (1995) 50
18. W.M. Fawley, "An Algorithm for Loading Shot Noise in Multi-Dimensional, Free-Electron Laser Simulation Code", to be published in Physical Review Special Topics –Accelerators and Beams
19. N.A.Vinokurov, "The Integral Equation for A High Gain FEL", ANL/APS/TB-27, 1996, Argonne National Laboratory
20. K.J.Kim, Phys. Rev. Lett. **57** (1986) 1871
21. S.Reiche et al., in Towards X-ray FELs (Garda Lake, Italy, 1997) AIP Conf. Proc. **413** (ed. R.Bonifacio and W.A.Barletta) (1997) 29
22. N.A.Vinokurov, private communication
23. L.D. Landau, E.M.Lifshitz, The Classical Theory of Field, Pergamon Press, 1975
24. V.Litvinenko, Physics of Super-pulses in storage ring FELs, this conference.
25. C.A.Brau, "Free-Electron Lasers", Academic Press, 1990
26. A. Murokh et al., submitted to Phys. Rev. Lett. (2002)